\documentclass[pra,a4paper,twocolumn,showpacs,superscriptaddress]{revtex4}
\usepackage{amsmath}
\usepackage{amsfonts}
\usepackage{graphicx}
\usepackage{longtable}
\begin{document}
\author{B. G. Divyamani } 
\affiliation{Department of Physics, Kuvempu University, Shankaraghatta, Shimoga-577 451, India.}
\affiliation{Tunga Mahavidyalaya, Tirthahalli, Shimoga-577 451,India}
\author{Sudha} 
\email{arss@rediffmail.com}
\affiliation{Department of Physics, Kuvempu University, Shankaraghatta, Shimoga-577 451, India.}
\affiliation{Inspire Institute Inc., Alexandria, Virginia, 22303, USA.}
\date{\today}
\title{Thermal entanglement in a two-qubit Ising chain subjected to Dzialoshinski-Moriya  interaction}   
\begin{abstract} 
Thermal entanglement of a two-qubit Ising chain subjected to an external magnetic field and Dzialoshinski-Moriya (DM) interaction is examined. The effect of magnetic field, strength of DM interaction and temperature are analyzed by adopting negativity of partial transpose as the measure of entanglement. It is shown that when the DM interaction along the Ising axis is considerable, thermal entanglement can be sustained for higher temperature. The usefulness of longitudinal DM interaction over the one that is perpendicular to the Ising axis, in the manipulation and control of entanglement at a feasible temperature, is illustrated. 
\end{abstract}
\maketitle
The study of thermal entanglement, the entanglement in the thermal equilibrium state of a quantum system,  particularly quantum spin chains in solid state systems, is known to provide a bridge between quantum information processing and condensed matter physics~\cite{Arnsen, xw1, xw1a,kamata,zhou,xw2,Gunlycke,chinphy,Buzek,Akyuz1,Dchuang,Ma,cpb,zhang,zhang2}. The usefulness of entangled spin chains in thermal equilibrium to the future realization of quantum computers~\cite{kane} necessitates this potentially rich area of study. In this context many authors have studied the entanglement properties in both the ground and thermal states of spin chains interacting through various Heisenberg interaction models such as $XX$, $XY$, $XXZ$ and $XYZ$~\cite{xw1,xw1a,kamata,zhou,xw2}. The study of thermal entanglement properties in solid state systems with Ising like interaction subjected to an external transverse magnetic field has also been carried out quite extensively~\cite{Gunlycke,chinphy,Buzek,Akyuz1}.  
 
Dzialoshinski-Moriya (DM) interaction~\cite{DM,Moriya}, an anisotropic and antisymmetric exchange interaction, arising due to spin orbit coupling is seen to enhance the thermal and ground state entanglement of Heisenberg spin chains~\cite{Dchuang,Ma,cpb,zhang}. The anisotropy and antisymmetry of the interaction is evident through its form $\stackrel{\rightarrow}{D}.[\stackrel{\rightarrow}{S_1}\times\stackrel{\rightarrow}{S_2}]$ ~\cite{DM, Moriya}. 
Thermal entanglement of a two-qutrit Ising system subjected to a magnetic field and DM interaction, both along the direction of Ising axis, is studied by C. Aky\"{u}z et. al.~\cite{Akyuz1}. Qin Meng et.al~\cite{cpb} have studied the effect of DM interaction on a two-qubit Heisenberg $XY$  spin chain with {\emph {transverse}} magnetic field and it reduces to the two-qubit Ising chain when the anisotropic parameter is $\pm 1$. Very recently, there has been an effort to analyze the thermal entanglement in a two-qubit Ising chain with inhomogeneous magnetic field and DM interaction both along the Ising axis~\cite{zhang2}. Despite these studies~\cite{cpb,zhang2} leading to interesting results, there has not been an explicit study on the effect of DM interaction on the thermal entanglement in a two-qubit Ising chain with either a transverse or a longitudinal magnetic field. The distinct roles of transverse or longitudinal magnetic fields in a two-qubit Ising chain with DM interaction is an interesting study in itself due to the importance of Ising interaction.  Towards this end, 
we examine the combined effect of external magnetic field as well as DM interaction on the variation of thermal entanglement in a two-qubit Ising chain. We analyze both the Ising model with longitudinal magnetic field, DM interaction and the transverse Ising model with DM interaction being perpendicular to the Ising direction and analyze the results.    

The one dimensional Ising model describes a set of linearly arranged spins, each interacting with its nearest neighbors by a coupling which is proportional to $\hat{\sigma}_{x}\otimes\hat{\sigma}_{x}$~\cite{Gunlycke}.
The two-qubit Ising chain subjected to magnetic field and DM interaction~\cite{DM} both along the direction of the Ising axis is modelled by the Hamiltonian
\begin{equation}
\label{model}
\hat{H} =2J (\hat{\sigma}_{1z}\cdot\hat{\sigma}_{2z})+B(\hat{\sigma}_{1z}+\hat{\sigma}_{2z}) + d (\hat{\sigma}_{1x}\cdot\hat{\sigma}_{2y}-\hat{\sigma}_{1y}\cdot\hat{\sigma}_{2x}).
\end{equation}
The first term  here represents the one dimensional Ising interaction and $\hat{\sigma}_{i\alpha},i=1,2(\alpha=x,y,z)$ are usual Pauli's spin operators. Here '$J$' denotes the coupling constant and the situations $J>0$, $J<0$ correspond respectively to the antiferromagnetic and ferromagnetic cases. The second term in Eq. (\ref{model}) corresponds to the external uniform magnetic field `$B$' to which the qubits are subjected to and '$d$' stands for DM interaction parameter.

The eigenvalues of the Hamiltonian $\hat{H}$ are seen to be 
\begin{eqnarray}
\lambda_1=2(J+B); & & \lambda_2=2(J-B); \nonumber \\
\lambda_3=-2(J-d); & & \lambda_4=-2(J+d); 
\end{eqnarray}
and the respective eigenvectors are 
\begin{eqnarray}
X_1=\vert 00\rangle; & &  X_3=\frac{\vert 01\rangle-i\vert 10\rangle}{\sqrt{2}}; \nonumber \\ 
X_2=\vert 11\rangle ; & & X_4=\frac{\vert 01\rangle+i\vert 10\rangle}{\sqrt{2}};
\end{eqnarray}
The eigenvectors $X_1$, $X_2$ belonging to the eigenvalues $\lambda_{1,\,2}$ (that contain magnetic field $B$) are separable while the eigenvectors $X_3$, $X_4$ belonging to $\lambda_{3,\,4}$ (that contain the DM interaction parameter $d$) are maximally entangled. Thus, it is readily seen that, without the DM interaction, there cannot be entanglement in the ground state of the system. 

The state of the physical system, described by the Hamiltonian $H$, at thermal equilibrium is given by $\rho(T)=\frac{e^{-\frac{H}{kT}}}{Z}$ where $Z=\mbox{Tr}\left[e^{-\frac{H}{kT}}\right]$ is the partition function and $k$ is Boltzmann's constant. The entanglement associated with the thermal state $\rho(T) $ is called thermal entanglement~\cite{Arnsen}. 
	
In order to quantify the thermal entanglement associated with $\rho(T)$, we have used the well known measure of entanglement namely negativity of partial transpose~\cite{vidal}.  The negativity $N(\rho)$ of a composite density matrix $\rho$ is equal to the sum of the absolute values of the negative eigen values of the partially transposed density matrix $\rho^\bot$ and it measures the degree to which $\rho^\bot$ fails to be positive~\cite{vidal}. For a bipartite system, the partial transpose~\cite{peres}, transposed with respect to the second subsystem is given by $(\rho^\bot)_{m\mu,n\nu}=\rho_{m\nu,n\mu}$, where the Latin indices refer to the first subsystem and the Greek indices refer to the second subsystem. It was shown in~\cite{Horodecki} that the positivity of the partial transpose(PPT) is a necessary and sufficient condition for separability of bipartite systems of dimensions $2 \times 2$ and $2 \times 3$. For higher dimensions the PPT criterion is only a necessary condition for separability. An unambiguous determination of $N(\rho)$ can be done using the relation $N(\rho)=\frac{||\rho^\bot||-1}{2}$, where $\vert\vert\rho^\bot\vert \vert$ is the trace norm of the partially transposed density matrix $\rho^\bot$. 

Having defined the model and the corresponding thermal state $\rho(T)$,	we now wish to investigate the entanglement in $\rho(T)$. The explicit form of the thermal density matrix $\rho(T)$ is given by
\begin{eqnarray}
\label{tdm}
\rho(T)=\left(\begin{array}{cccc}m_{1}&0&0& 0 \\0& m_{2}&-i n_{2}&0\\0& i n_{2} & m_{2} &0\\ 0 &0&0& m_{4}\end{array}\right) \ \mbox{where} \ 
\end {eqnarray}
\begin{eqnarray}
\label{mn1}
m_{1}&=& \frac{1}{1+e^{\frac{4B}{T}}+2e^{\frac{2(2J+B)}{T}}\cosh{\frac{2d}{T}}};  \nonumber \\
m_{2}&=& \frac{e^{\frac{2(2J+B)}{T}}\cosh{\frac{2d}{T}}}{1+e^{\frac{4B}{T}}+ 2e^{\frac{2(2J+B)}{T}}\cosh{\frac{2d}{T}}}; \nonumber  \\
n_{2}&=& \frac{e^{\frac{2(2J+B)}{T}}\sinh{\frac{2d}{T}}}{1+e^{\frac{4B}{T}}+ 2e^{\frac{2(2J+B)}{T}}\cosh{\frac{2d}{T}}};  \nonumber \\
m_{4}&=&\frac{e^{\frac{4B}{T}}}{1+e^{\frac{4B}{T}}+ 2e^{\frac{2(2J+B)}{T}}\cosh{\frac{2d}{T}}};  
\end{eqnarray}

We will now examine the dependence of negativity of partial transpose $N(\rho)$ of the thermal density matrix $\rho(T)$
on the associated parameters $J$, $B$, $d$ and $T$. As the thermal density matrix $\rho(T)$ is of the form given in Eq (\ref{tdm}), it can be readily seen that the partially transposed density matrix $\rho(T)^\bot$, transposed with respect to the second qubit, is given by  
\[
\rho(T)^\bot=\left(\begin{array}{cccc}m_{1}&0&0&-in_{2}\\0&m_{2}&0 &0\\0& 0 &m_{2}&0\\i n_{2} & 0 & 0 &m_{4}\end{array}\right)
\]
where $m_i$ and $n_j$ are given in Eq. (\ref{mn1}). The square root of the eigen values of  $\rho(T)\rho(T)^\bot$  are given by 
\begin{small}
\begin{eqnarray}
\mu_{1,2}&=& m_2  \nonumber  \\
\mu_{3,4}&=&\sqrt{\frac{m_1^2+m_4^2+2n_2^2 \pm (m_1+m_4) \sqrt{(m_{1}-m_{4})^2+4n^2_{2}}}{2}}\nonumber 
\end{eqnarray}
\end{small}
and we have the negativity of partial transpose $N(\rho)$ to be  
\begin{eqnarray}
\label{Nt}
N(\rho)=\frac{\left(\sum_{i=1}^4\,\mu_i\right)-1}{2}.
\end{eqnarray}
The maximum value of $N(\rho)$ for a two-qubit state is $1/2$.

The following graphs are effective to capture the dependence of $N(\rho)$ on the physical parameters $T$, $B$ and $d$. 
\begin{figure}[ht]
\includegraphics*[width=2in,keepaspectratio]{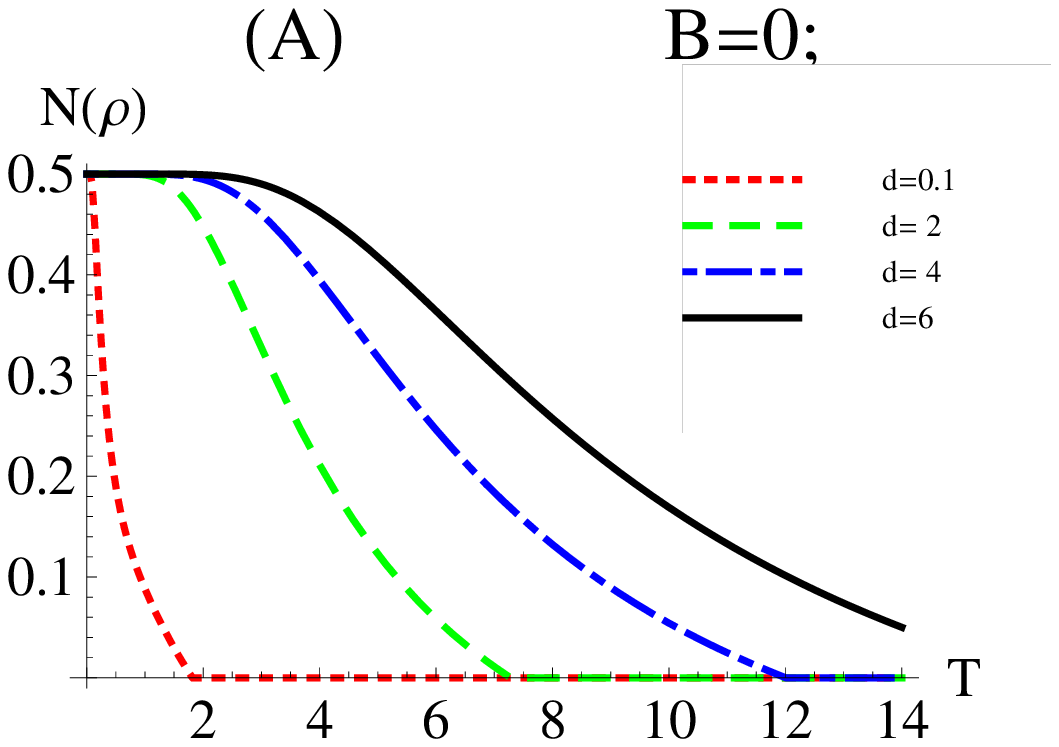} 
\includegraphics*[width=2in,keepaspectratio]{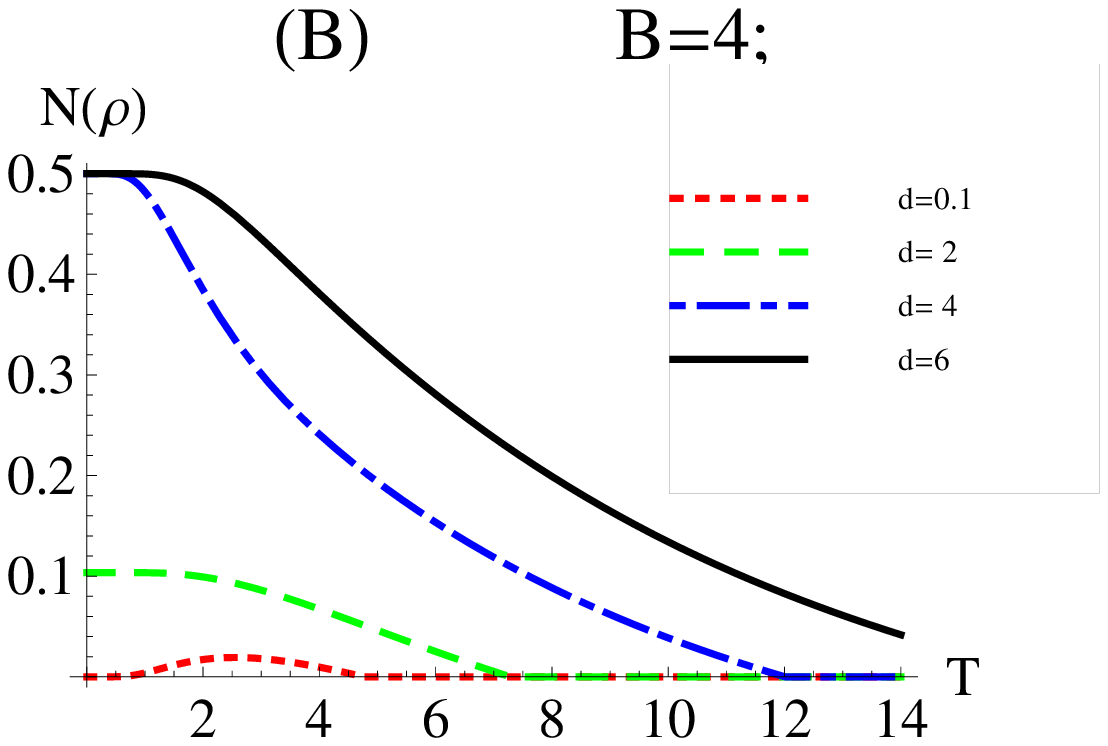}
\caption{
Two dimensional plots showing the effect of magnetic field on $N(\rho)$. ($J=1$.)}
\end{figure} 
\begin{figure}[ht]
\includegraphics*[width=2in,keepaspectratio]{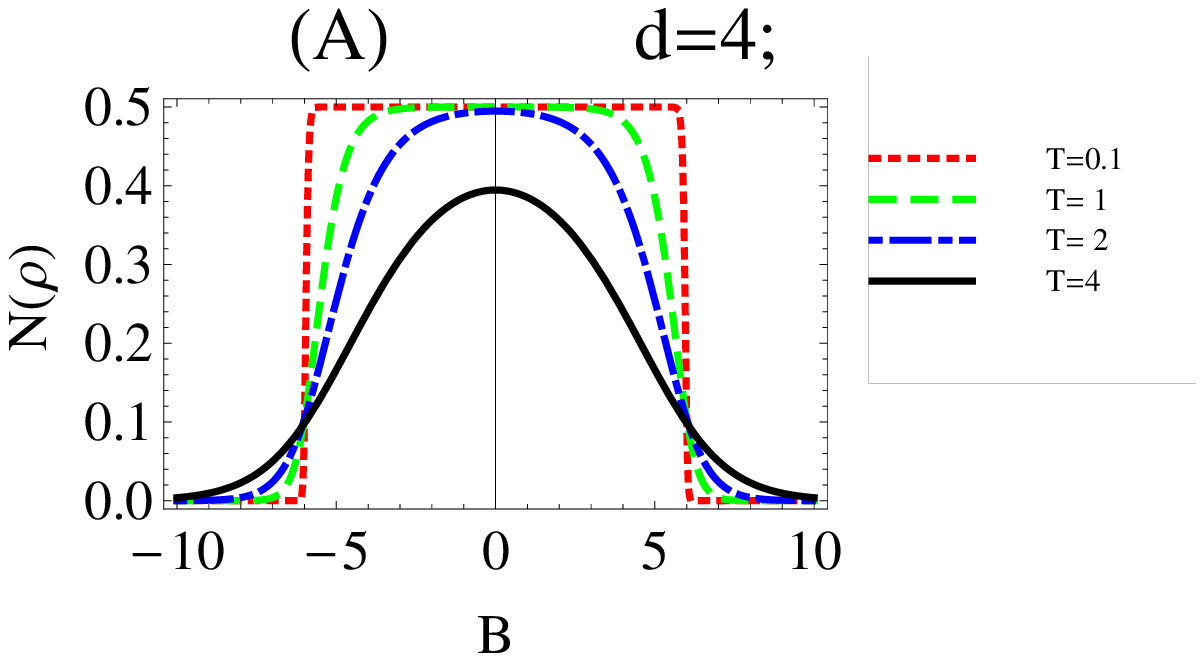}
\includegraphics*[width=2in,keepaspectratio]{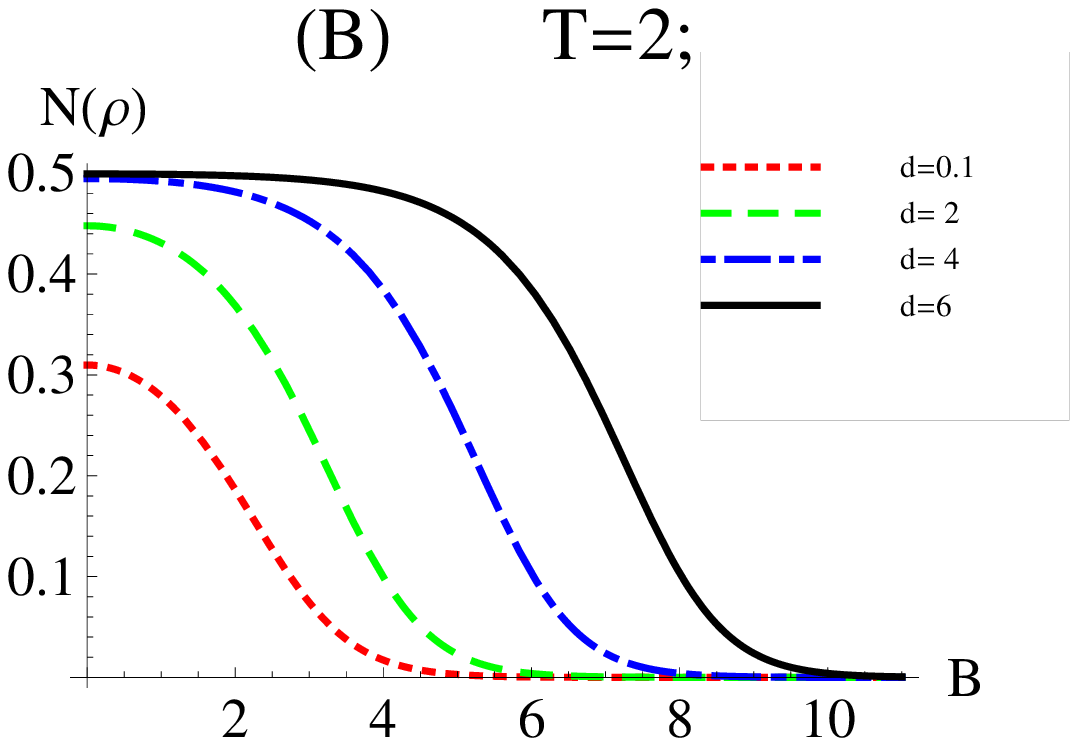}  
\caption{
Variation of $N(\rho)$ with magnetic field for fixed $d$ and $T$ ($J=1$.)    }
\end{figure}
\begin{figure}[ht]
\includegraphics*[width=2.2in,keepaspectratio]{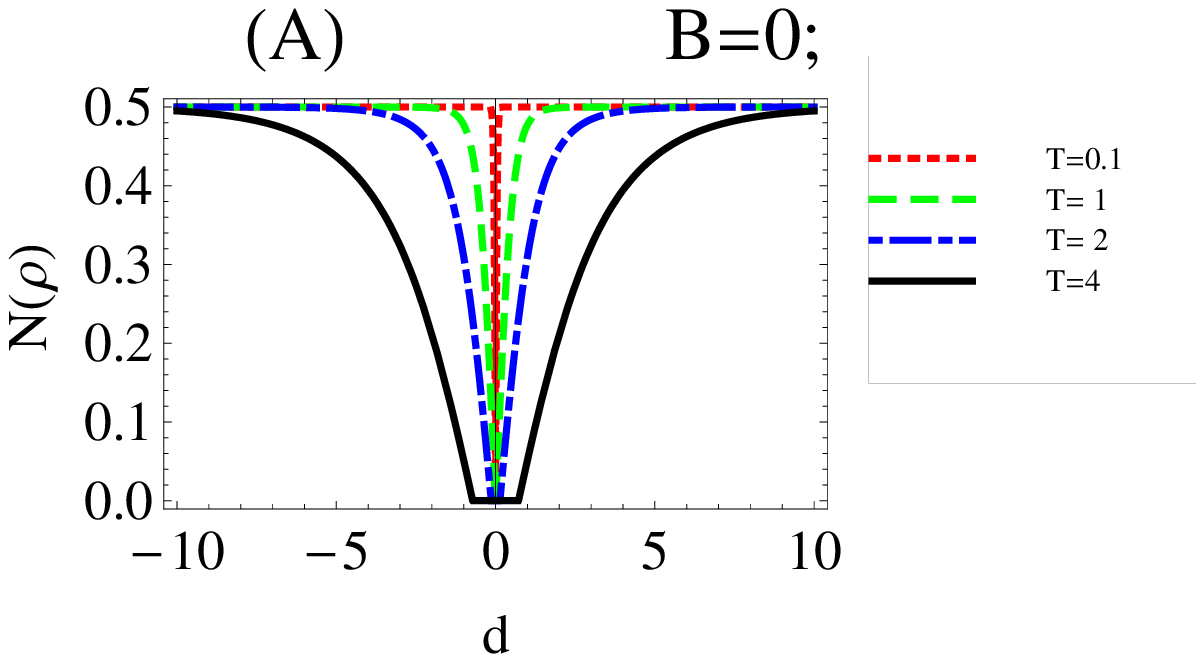}
\includegraphics*[width=2.2in,keepaspectratio]{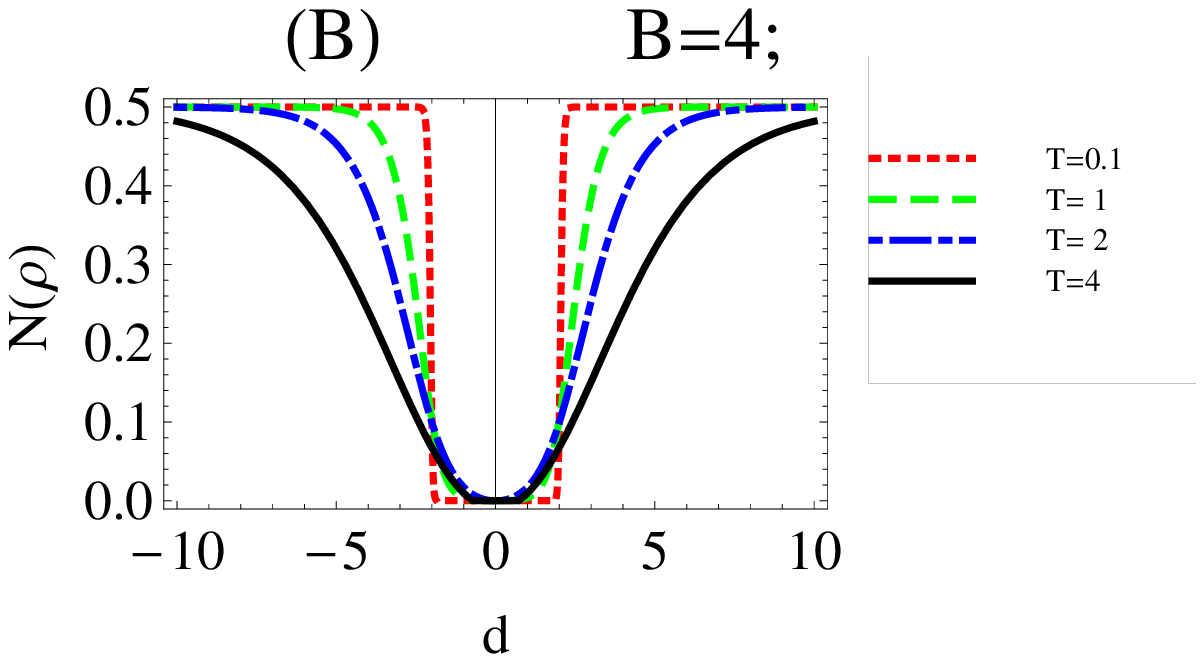}  
\caption{
Variation of $N(\rho)$ with DM interaction parameter $d$ for fixed magnetic field $B$ at different values of $T$ ($J=1$.)}
\end{figure}
It can be seen through Figs. (1)--(3) that
\begin{itemize} 
\item[(1)] An infinitesimal DM interaction also causes maximum entanglement at $T=0$ and $B=0$. The destructive role played by the longitudinal magnetic field on $N(\rho)$ is also readily seen(See Figs. 1(A), 1(B), 3(A), 3(B)).  
\item[(2)]  The range of temperature over which the thermal entanglement persists increases with the increase in DM interaction (See Fig. 1(A)).   
\item[(3)] For any fixed non-zero value of $d$, there exists a critical magnetic field above which the entanglement vanishes. At $T\approx0$, the vanishing of $N(\rho)$ is quite sudden at the critical magnetic field $B_0$ but for higher values of $T$, $N(\rho)$ decreases smoothly and even extend beyond the critical magnetic field $B_0$(See Fig. 2(A)). 
\end{itemize} 
Instead of the longitudinal magnetic field, if we consider a transverse external magnetic field and the DM interaction perpendicular to the Ising interaction, the Hamiltonian of the system is given by 
\begin{equation}
\label{model2}
\hat{H} =2J (\hat{\sigma}_{1x}\cdot\hat{\sigma}_{2x})+B(\hat{\sigma}_{1z}+\hat{\sigma}_{2z}) + d (\hat{\sigma}_{1x}\cdot\hat{\sigma}_{2y}-\hat{\sigma}_{1y}\cdot\hat{\sigma}_{2x}).
\end{equation}
Here we have chosen the Ising axis to be the $x$-axis and the magnetic field, DM interaction along z-axis.  
The eigenvectors of $\hat{H}$ are seen to be 
\begin{eqnarray}
X_1&=&J\vert 00\rangle+\left(\sqrt{J^2+B^2}-B\right)\vert 11\rangle\in 2\sqrt{J^2+B^2} \nonumber \\ 
X_2&=&\left(B-\sqrt{J^2+B^2}\right)\vert 00\rangle+J \vert 11\rangle\in -2\sqrt{J^2+B^2} \nonumber \\
X_3&=&(J-id)\vert 01\rangle+\sqrt{J^2+d^2}\,\vert 10 \rangle \in 2\sqrt{J^2+d^2}  \\
X_4&=&\sqrt{J^2+d^2}\,\vert 01 \rangle-(J-id)\vert 10\rangle \in -2\sqrt{J^2+d^2}.\nonumber
\end{eqnarray}
Here two of the eigenvalues depend on $J$ and $B$ while the other two depend on $J$, $d$. 
All the eigenvectors are entangled for non-zero values of $B$ and $d$. This indicates that the ground state of the system is entangled when either $B$ or $d$ or both of them are non-zero. In order to see the combined effect of $B$ and $d$, we evaluate the thermal density matrix and the negativity of partial transpose, in the same manner as is done  for the case of longitudinal magnetic field. Choosing to denote the negativity of partial transpose here as $N^\bot(\rho)$ ($'\bot'$ denoting the transversality of magnetic field and DM interaction). The variation of $N^\bot(\rho)$ with the parameters $T$, $B$ and $d$ are as shown in the following figures. 
\begin{figure}[ht]
\includegraphics*[width=2in,keepaspectratio]{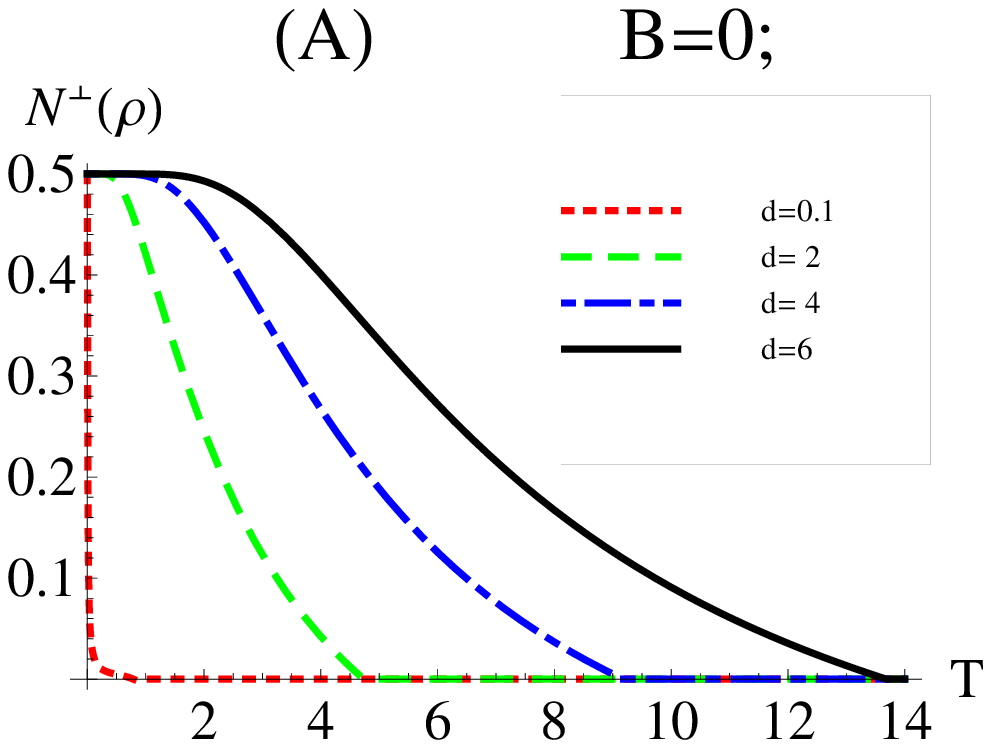} 
\includegraphics*[width=2in,keepaspectratio]{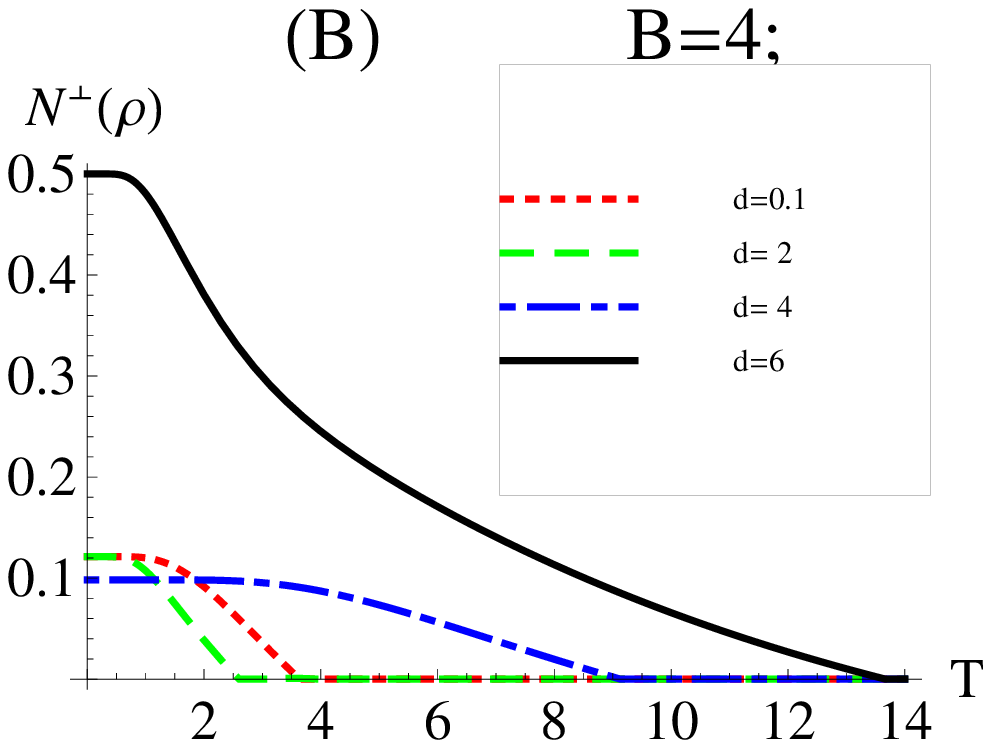}
\caption{
Two dimensional plots showing the effect of transverse magnetic field on $N^\bot(\rho)$. ($J=1$.)}
\end{figure} 
\begin{figure}[ht]
\includegraphics*[width=2in,keepaspectratio]{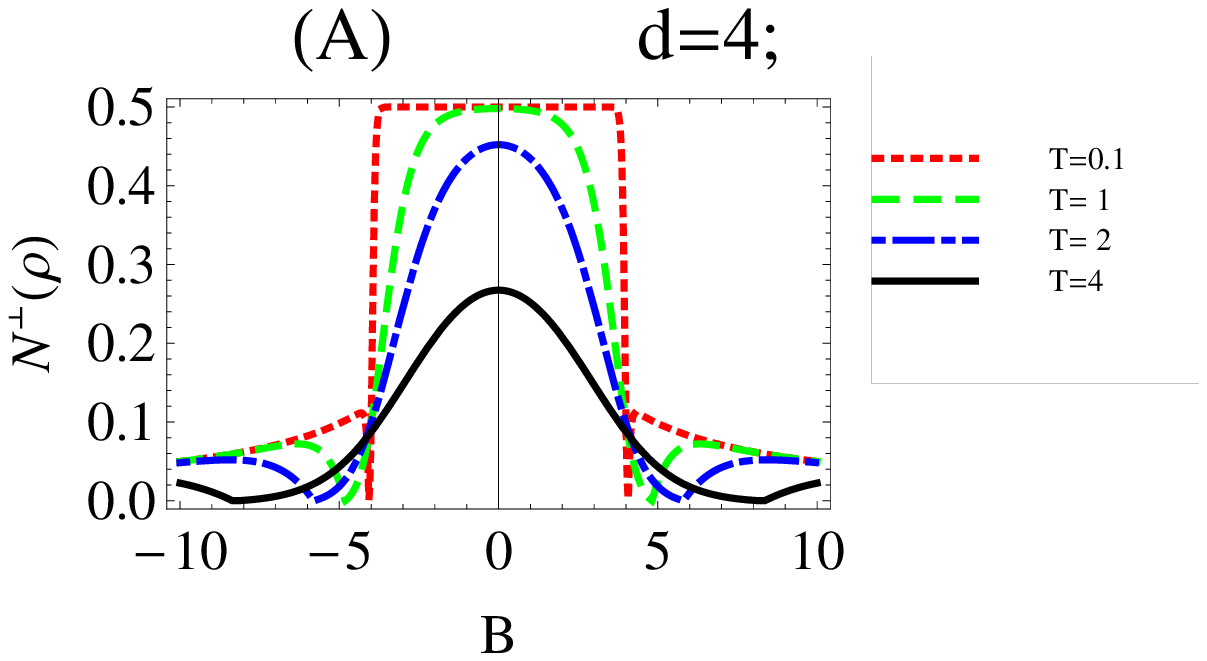}
\includegraphics*[width=2in,keepaspectratio]{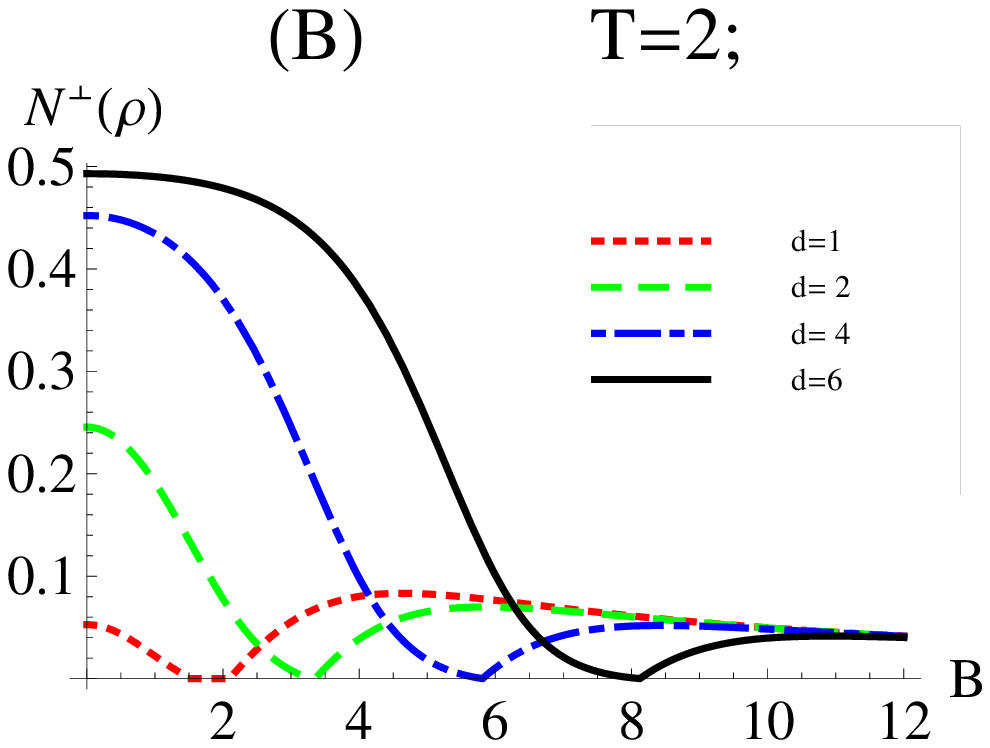}  
\caption{
Variation of $N^\bot(\rho)$ with transverse magnetic field for fixed $d$ and $T$ ($J=1$.)    }
\end{figure} 
\begin{figure}[ht]
\includegraphics*[width=2.2in,keepaspectratio]{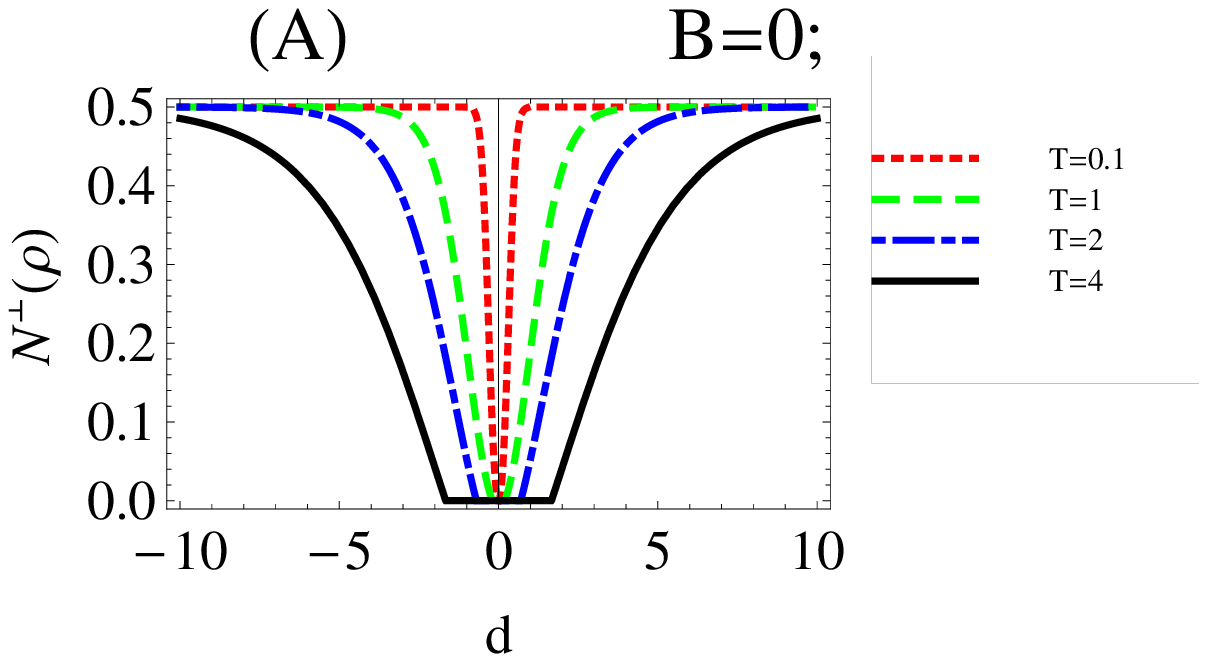}
\includegraphics*[width=2.2in,keepaspectratio]{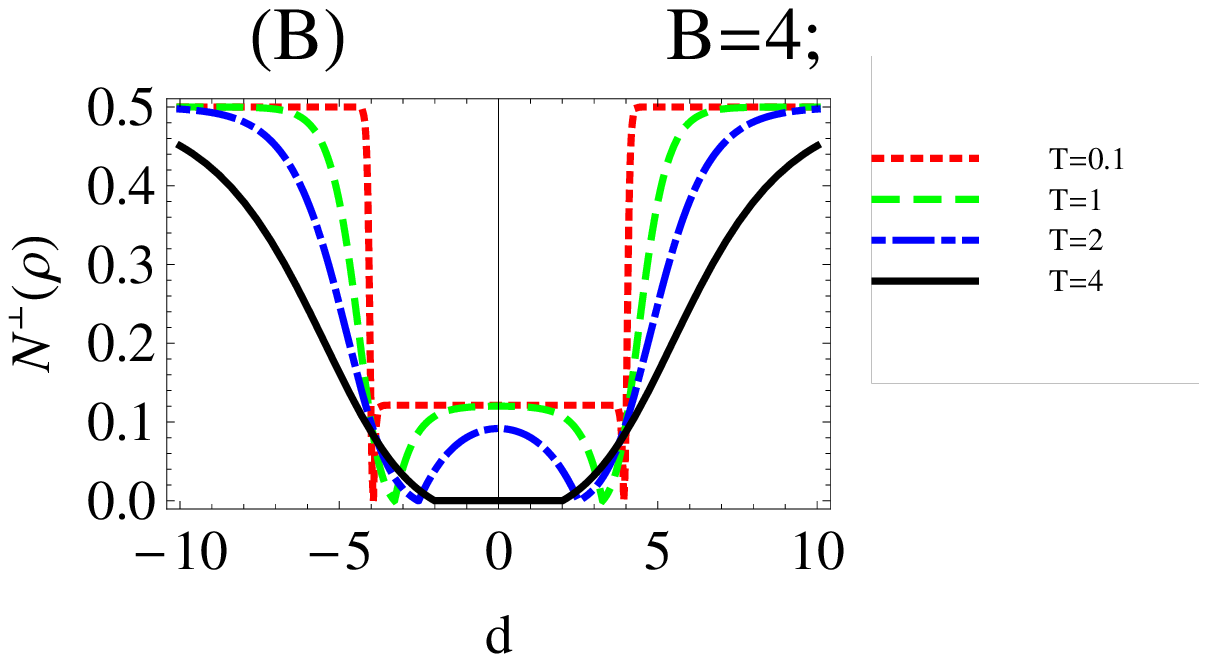}  
\caption{
Variation of $N^\bot(\rho)$ with DM interaction parameter $d$ for fixed magnetic field $B$ at different values of $T$ ($J=1$.)}
\end{figure}

An interesting feature in the case of transverse magnetic field and DM interaction is the vanishing and recovery of entanglement at a particular value of magnetic field, that depends both on temperature and strength of DM interaction (See Figs. 5(A) and 5(B)). In particular, at $T=0$, $N^\bot(\rho)$ suddenly drops to zero from its maximum value at $B=d$. For $T>0$, the vanishing and recovery  are not-so-sudden and they  occur at $B\approx d+T$. 

It is to be noticed here that at any temperature there is a threshold value of transverse magnetic field $B$ that helps (considerably so at low enough temperatures) the thermal entanglement  and a further increase in $B$ will reduce the thermal entanglement. In contrast, an increase in the DM interaction (both transverse and longitudinal) always aids the thermal entanglement when there is no magnetic field. But an increase in the magnetic field reduces the effect of DM interaction (See Figs. 1 and 4) and hence the thermal entanglement vanishes at $B\approx d+T$. Even when the effect of DM interaction is nearly nullified due to magnetic field, the thermal entanglement still persists due to the presence of transverse magnetic field. That is, the thermal entanglement after vanishing recovers back to its value that is due to transverse magnetic field alone. (Such a recovery does not happen in the case of longitudinal magnetic field because longitudinal magnetic field does not assist thermal entanglement). Fig. 7 below illustrates our point.     
\begin{figure}[ht]
\includegraphics*[width=2.4in,keepaspectratio]{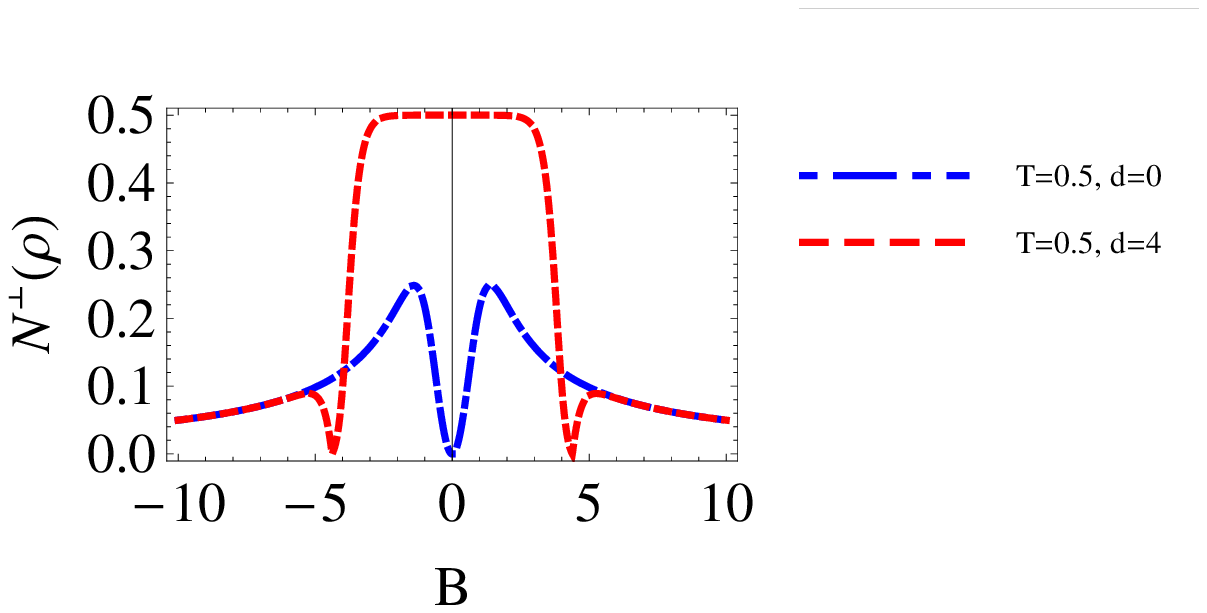}
\caption{
The competing behaviours of transverse magnetic field and DM interaction  ($J=1$.)}
\end{figure}

In both the models that we have considered, the sign of the parameter $d$ and that of the magnetic field $B$ do not affect the thermal entanglement. 
While the thermal entanglement in transverse Ising model is known to be symmetric over the magnetic field $B$~\cite{Gunlycke}, the thermal entanglement in the presence of DM interaction in Heisenberg XY model is shown to be independent of the sign of the DM interaction parameter $d$~\cite{cpb}. Thus, it is not surprising that the thermal entanglement in the more simpler Ising models that we have examined here show symmetry in the DM interaction parameter $d$, in addition to that of magnetic field $B$.

The role of transverse magnetic field in being able to create thermal entanglement at lower temperatures, even without DM interaction, can be readily seen through Fig. 6(B). Still, DM interaction is more beneficial for thermal entanglement at higher temperatures and the transverse magnetic field plays a destructive role, quite similar to the longitudinal field, at higher temperatures. While a longitudinal magnetic field cannot create thermal entanglement without DM interaction, a transverse magnetic field can do so. When there is no magnetic field, a longitudinal DM interaction is seen to be more helpful than the transverse DM interaction in the creation and control of thermal entanglement (See Figs. 1(A), 3(A), 4(A) and 6(A)).  Thus we can conclude that a pure longitudinal DM interaction, without either a transverse or longitudinal magnetic field, helps in creating and sustaining thermal entanglement at reasonably high temperatures.

In this article, we have examined the nature of variation of thermal entanglement of a two-qubit Ising chain kept in a magnetic field and subjected to Dzialoshinski-Moriya (DM) interaction arising due to spin-orbit interaction. The situation in which the magnetic field and DM interaction are along the Ising direction is analyzed. It is shown that a pure DM interaction (without  magnetic field) along the Ising axis can give rise to a thermal entanglement and a larger value of the DM interaction parameter is shown to result in a larger range of temperature over which the entanglement persists.  We have also analyzed the situation in which the Ising chain is subjected to a magnetic field and DM interaction, both being perpendicular to the Ising direction. 
The longitudinal DM interaction is found to be more beneficial than the transverse DM interaction in sustaining the thermal entanglement over a larger range of temperature. Both the longitudinal and transverse magnetic fields do not seem to aid the thermal entanglement at higher temperatures and a pure longitudinal DM interaction is thus seen to be the right option for thermal entanglement in one-dimensional Ising system with qubits. 
We conjecture that these results are applicable to the pairwise entanglement in an $N$-qubit Ising chain in the presence of DM interaction.      

\end{document}